\documentclass[final,3p,times,authoryear]{elsarticle}

\usepackage{amssymb}
\usepackage{amsmath}

\usepackage{graphicx}
\usepackage{graphics}
\usepackage{dcolumn}
\usepackage{color}
\usepackage{rotate}
\usepackage{fancyhdr} 
\usepackage{ulem}

\long\def\comment#1{}
\def\be{\begin{equation}}
\def\ee{\end{equation}}
\def\bea{\begin{eqnarray}}
\def\eea{\end{eqnarray}}
\def\ba{\begin{eqnarray}}
\def\ea{\end{eqnarray}}

\def\VEV#1{\left\langle #1 \right\rangle}

\definecolor{darkred}{rgb}{.743,0,0}

\journal{Annals of Physics}

\begin{document}

\begin{frontmatter}

\title{Cosmological-perturbation equations from the total-angular-momentum formalism}

\author[first]{Marc Kamionkowski}
\affiliation[first]{organization={William H. Miller III Department of Physics and Astronomy},
            addressline={Johns Hopkins University}, 
            city={Baltimore},
            state={MD},
            postcode={21218}, 
            country={USA}}
\author[second]{Robert R.\ Caldwell}    
\affiliation[second]{organization={Department of Physics and Astronomy},
            addressline={Dartmouth College}, 
            city={Hanover},
            state={NH},
            postcode={03755}, 
            country={USA}}

\date{\today}

\begin{abstract}
We derive the equations for the linear evolution of cosmological perturbations using total-angular-momentum (TAM) waves.  The CMB temperature and polarization transfer functions are first derived as integral equations and from these the usual Boltzmann hierarchy of differential equations are then obtained for both scalar and tensor perturbations.  We include the effects of isotropic cosmic  birefringence and discuss anisotropic cosmic birefringence.  We note a subtle effect of cosmic birefringence on temperature fluctuations.  The TAM derivation of the equations provides physical insights to results obtained previously.
\end{abstract}

\end{frontmatter}


\section{Introduction}

Work on the linear evolution of cosmological perturbations traces back to the middle of the last century
\citep{Lifshitz:1945du,Silk:1967aha,Sunyaev:1970er,   Peebles:1970ag}.  Most modern approaches to the calculation of the key observables---the matter power spectrum and the cosmic microwave background (CMB) temperature and polarization power spectra---owe much to work in the
1980s \citep{Wilson:1981yi,Bond:1984fp,Bond:1987ub,Vittorio:1984aaz} in which the time evolution of the
fluctuations was quantified through the solution of a set
of coupled differential equations for the Einstein equations,
dark matter, baryon, neutrino, and photon density and velocity
perturbations, and the higher moments of the momentum
distribution functions for photons and neutrinos.  These equations have been elaborated and reviewed many times \citep{Bardeen:1980kt,Kodama:1984ziu,Mukhanov:1990me,Liddle:1993fq,Ma:1995ey,Malik:2008im,Kandus:2010nw} and are now part of the standard cosmology curriculum \citep{Mukhanov:2005sc,Durrer:2008eom,Dodelson:2020bqr,Baumann:2022mni,Huterer:2023mmv}. 

In this paper, we will re-derive these equations once again, but with some differences and new ingredients that may provide insights and possibly even practical value.  The primary novelty is the use of the total-angular-momentum \citep{Dai:2012bc} formalism (see also \cite{Challinor:1999xz} for a related approach).  In all past work, the equations have been derived by first decomposing  perturbation variables into Fourier modes, or plane waves.  Plane waves are used as they provide a complete orthonormal set for any possible perturbation, and they are simple and universally understood.  Equations for the metric perturbations, densities, velocities, and higher moments are then obtained for an individual Fourier mode.  At the end of the calculation, those plane waves are then projected onto spherical-sky observables.

With the approach presented here, perturbations are decomposed into total-angular-momentum (TAM) waves---eigenstates of the Helmholtz equation as well as eigenstates of rotations---that provide an alternative complete orthonormal set of basis functions.   With this approach, the symmetries of the celestial sphere are incorporated from the very start.  Plane waves are eigenstates, with vector-valued eigenvalue ${\bf k}$, of the spatial-translation operator, as well as eigenstates of eigenvalue $k^2$ of the Helmholtz equation.  TAM waves are also eigenstates of the Helmholtz equation but are instead eigenstates of eigenvalue $lm$ of rotation operators.  With TAM waves, any properly defined spherical-harmonic observable (like   $T_{lm}$, $E_{lm}$, and $B_{lm}$, for CMB temperature and polarization or $a_{lm}^\phi$ for the lensing potential ) receives contributions {\it only from TAM waves with the same} $lm$.\footnote{This approach has similarity to that in \cite{Hu:1997hp}, although there, the decomposition was still in terms of plane waves.}

This approach is also relevant for integral-equation formalisms for perturbations.  It has long been understood that the ``Boltzmann hierarchy,'' the infinite tower of differential equations for the evolution of the higher moments of the photon and neutrino distribution functions, can be replaced with integral equations \citep{Seljak:1996is,Weinberg:2003ur,Baskaran:2006qs,  Weinberg:2006hh,Kamionkowski:2021njk}.  Recently, this approach has been used \citep{Ji:2022iji,Lee:2025zym,Lee:2025vgv} to develop a new code, {\tt CLASSIER}, that replaces the time-consuming massive-neutrino Boltzmann hierarchies in {\tt CLASS} \citep{Lesgourgues:2011rh} with far more efficient integral-equation solutions.  And even more recently, it has been used to develop a fast code {\tt CLASSIER-DDM} \citep{Bencke:2026uws,Nanoominprep} for models with decaying dark matter.  In this paper, we derive some of the equations that were stated without derivation in \cite{Kamionkowski:2021njk} and \cite{Ji:2022iji}.

We also include cosmic birefringence \citep{Harari:1992ea,Carroll:1989vb,Lue:1998mq,Lepora:1998ix,Carroll:1998zi}, the rotation of the linear polarization in the presence of a time-evolving scalar field that has a parity-breaking coupling to electromagnetism.  Prior work \citep{Liu:2006uh,Murai:2022zur,Finelli:2008jv,Gubitosi:2014cua} provided generalizations to the Boltzmann and integral equations to include cosmic birefringence, but the derivation here provides a cross-check and perhaps new insights.  We note, for example, a subtle effect of cosmic birefringence on the temperature fluctuation.  We also consider in an Appendix anisotropic birefringence, but only to linear order in the rotation angle. The cosmic-birefringence aspect of this work is particularly timely given the case made recently \citep{Namikawa:2025doa} that cosmic birefringence may allow a higher optical depth and thus relax neutrino-mass tensions with DESI data.

This paper is organized as follows.  Section \ref{sec:tamtools} provides an overview of the calculation and reviews some of the basics of TAM waves and also some TAM relations used in the calculation.  Section \ref{sec:boltzmann} reviews the Boltzmann equation for the time evolution of the photon distribution function.  Section \ref{sec:scalar} then provides the central results of this paper: a derivation, using the TAM formalism, of integral-equation expressions for the temperature and polarization fluctuations induced by scalar metric perturbations.  Section \ref{sec:cb} explains how those results are augmented if there is cosmic birefringence.  Section \ref{sec:tensors} then derives results for temperature and polarization fluctuations for primordial tensor modes (gravitational waves). Section \ref{sec:conclusion} provides closing remarks.   \ref{app:projections} provides useful projections of TAM waves.  \ref{app:scattering} discusses the Thomson-scattering term in the Boltzmann equation. \ref{app:boltzmann} recovers the Boltzmann hierarchies from the integral equations.  And \ref{app:aniso} treats anistropic cosmic birefringence.

\section{Overview and Total Angular Momentum Waves}
\label{sec:tamtools}

\subsection{CMB observables}

The CMB temperature $T(\vec x,\hat n,\tau)$ and polarization ${\cal P}_{ab}(\vec x,\hat n,\tau)$ seen in a direction $\hat n$ by an observer at comoving position $\vec x$ and conformal time $\tau$ can be expanded,\footnote{In order to reduce notational clutter, throughout this paper, our ``temperature'' $T$ is actually the (dimensionless) fractional temperature perturbation and the polarization ${\cal P}_{ab}$ and Stokes parameter $Q$ and $U$ in units of the mean temperature.}
\begin{equation}
     T(\vec x,\hat n, \tau) =\sum_{lm} T_{lm}(\vec x,\tau) Y_{(lm)}(\hat n), \qquad
    {\cal P}_{ab}(\vec x, \hat n, \tau) = \sum_{lm} \left[ E_{lm}(\vec x,\tau) Y_{(lm)ab}^{\rm E}(\hat n) +  B_{lm}(\vec x,\tau) Y_{(lm)ab}^{\rm
     B}(\hat n) \right],
\label{eqn:TPexpansions}     
\end{equation}
at each spacetime point
in terms of their spherical-harmonic coefficients $T_{lm}(\vec x,\tau)$, $E_{lm}(\vec x,\tau)$, and $B_{lm}(\vec x,\tau)$. Here $Y_{lm}(\hat n)$ are the usual spherical harmonics and $Y_{lm(ab)}^E(\hat n)$ and $Y_{lm(ab)}^B(\hat n)$ tensor spherical harmonics (see, e.g., \cite{Kamionkowski:2015yta}).  The central aim of this paper will be integral equations for the observables that we see at our position $\vec x=0$ at the current conformal time $\tau_0$.

\subsection{Total angular momentum waves}

We will proceed by expanding perturbations in terms of total angular momentum (TAM) waves \citep{Dai:2012bc}, rather than plane waves, as is usually done. Like Fourier modes, TAM waves provide complete orthonormal bases for scalars, vectors, and tensors on three-dimensional Euclidean space.  As long as the fluctuation amplitude is small, each of these modes then evolves independently.  The problem then boils down to the calculation of the $T_{lm}$, $E_{lm}$, and $B_{lm}$ that arise from any given TAM wave.  As shown in \cite{Dai:2012bc}, for observables of a given $lm$, only TAM waves of the same $lm$ contribute.

We will summarize properties of scalar and tensor total-angular-momentum (TAM) waves \citep{Dai:2012bc} that we will use.  TAM waves are solutions of the Helmholtz equation, $(\nabla^2+k^2)\Psi=0$, of fixed total angular momentum $lm$.  The scalar TAM waves---also known as Fourier-Bessel modes \citep{Heavens:1994iq,Heavens:2003jx,Leistedt:2011mk}---and their orthonormality relation are
\begin{equation}
     \Psi_{klm}(\vec x) = j_l(kx) Y_{(lm)}(\hat x), \qquad       \int\, d^3 x\, \left[ 4\pi i^l\Psi_{klm}(\vec x) \right]^* \left[4 \pi i^{l'} \Psi_{k'l'm'}(\vec x) \right] = \delta_{kk'} \delta_{ll'}\delta_{mm'},
\label{eqn:scalarTAM}
\end{equation}
where $\delta_{kk'}$ is shorthand for $(2\pi)^3\delta_D(k-k')/k^2$.
Any scalar function may thus be expanded as
\begin{equation}
     X(\vec x) = \sum_{klm} 4 \pi i^l X_{klm}(\vec x) \Psi_{klm}(\vec x), \qquad
    {\rm with} \qquad X_{klm}= \int\, d^3x \, \left[ 4\pi i^l \Psi_{klm}(\vec x) \right]^* X(\vec x) = \int d^2 \hat k \,\tilde X(\vec k) Y_{(lm)}^*(\hat k),
\label{eqn:TAMexpansions}    
\end{equation}
with $\sum_k \equiv \int k^2\,dk/(2\pi)^3$ and
where the last relation provides TAM coefficients $X_{klm}$ in terms of the Fourier amplitudes $\tilde X(\vec k) = \int d^3 x\,X(\vec x) e^{-i \vec k \cdot \vec x}$.  If $X(\vec x)$ is a random field characterized by $\VEV{ \tilde X(\vec k) \tilde X^*(\vec k')} = (2\pi)^3 \delta_D(\vec k-\vec k') P_X(k)$, in terms of a power spectrum $P_X(k)$, then the TAM coefficients are random variables that satisfy $\VEV{X_{klm} X^*_{k'l'm'}}=\delta_{ll'}\delta_{mm'}\delta_{kk'} P_X(k)$.  

The most general symmetric tensor can be expanded as 
\begin{equation}
    T_{ab}(\vec x)= \sum_\alpha \sum_{klm} 4 \pi i^l T^\alpha_{klm} \Psi^{\alpha}_{(klm)ab}(\vec x).
\end{equation}
Here, $\alpha$ sums over the 5 traceless components which include a longitudinal (L), E-mode and B-mode vector (VE and VB), and a transverse traceless E-mode and B-mode (TE and TB), as well as a sixth mode $\Psi^{t}_{(klm)ab}(\vec x) = (1/3)\Psi_{(klm)}(\vec x) g_{ab}$ to account for the trace.  In this paper we will need the trace and longitudinal mode for scalar perturbations and the TE and TB modes for tensor perturbations.

The orthonormality relations of the  five TAM wave modes are given by \citep{Dai:2012bc},
\begin{equation}
    (4\pi)^2 \int d^3 x \left[ \Psi^{\, \alpha}_{(klm)ab} ({\bf x})\right]^* \, \Psi^{\,\beta,\, ab}_{(k'l'm')}({\bf x}) = \delta_{\alpha\beta}\, \delta_{ll'} \delta_{mm'} \delta_{kk'},
\end{equation}
where $\{\alpha,\beta\} = \{L, VB, VE, TB, TE\}$.
A tensor-valued plane wave can be expanded,
\begin{equation}
     \varepsilon_{ab}(\vec k)  e^{i \vec k \cdot \vec x} = \sum_\alpha \sum_{lm} 4 \pi i^l B^\alpha_{lm}(\hat k) \Psi^\alpha_{(klm)ab}(\vec x), \qquad  \mathrm{with} \qquad B^\alpha_{lm}(\hat h) = \varepsilon^{ab}(\vec k) Y^{\alpha\,*}_{(lm)ab}(\hat k),
\end{equation}
in terms of tensor TAM waves.

Below we deal with cosmological perturbation variables $X(\vec x,\tau)$ that depend on conformal time $\tau$ and position $\vec x$.  Given rotational invariance, the time evolution of the TAM-wave amplitudes is the same for all $lm$ of the same $k$ (analogous to the fact that in the Fourier analysis, the time evolution depends only on the magnitude $k$ of the Fourier wavevector $\vec k$, and not its direction).  Thus, it will be possible to write these $X_{klm}(\vec x,\tau)=A_{klm} X_k(\tau)$ in terms of amplitudes $A_{klm}$ and $lm$-independent time dependence $X_k(\tau)$.  These time dependences are then the same as in the Fourier approach.

If we have scalar perturbations, then vector-valued quantities (like the peculiar velocity) and tensor-valued quantities (like the photon intensity quadrupole) are constructed from the longitudinal vector and tensor TAM waves, and for the tensor-valued quantities, also the trace.
These longitudinal vector and tensor TAM waves of the same $lm$ are obtained by applying derivative operators to $\Psi_{klm}(\vec x)$,
\begin{equation}
     \Psi^L_{(klm)a}(\vec x) = \frac{i}{k}\nabla_a \Psi_{klm}(\vec x),
     \qquad
     \Psi^L_{(klm)ab}(\vec x) = -\sqrt{\frac32} \frac{1}{\nabla^2}
     \left(\nabla_a\nabla_b - \frac13 g_{ab}\nabla^2\right)\Psi_{klm}(\vec x) \equiv - W^L_{ab} \Psi_{klm}(\vec x),
\label{eqn:longwaves}
\end{equation}
which are the unique spatial dependences that a velocity field and a
quadrupole field sourced by the scalar mode
$\Psi_{klm}$ may have.  Throughout, $g_{ab}$ is the metric on flat three-dimensional space, and we have defined the differential operator $W^L_{ab}$ in the last equality.\footnote{Note that our $W^L_{ab}$ is not the same as that in \cite{Dai:2012bc}.}

For tensor perturbations we need the two
transverse-traceless waves $\Psi^{TE}_{(klm)ab}(\vec x)$ and $\Psi^{TB}_{(klm)ab}(\vec x)$,
of parity $(-1)^l$ and $(-1)^{l+1}$, respectively, defined in Eqs.~(74) and (85) of \cite{Dai:2012bc}.  We will need the projections of the tensor TAM waves on the line-of-sight and transverse directions.  These are obtained from Eq. (94) in \cite{Dai:2012bc}, and the results are summarized in \ref{app:projections}.

\section{Boltzmann equation for photon distribution function}
\label{sec:boltzmann}

\subsection{Preliminaries}

We surmise that the Universe is described by the metric
\begin{equation}
     ds^2 = a^2(\tau) \left[ -d\tau^2 + \left(g_{ab}+h_{ab}(\vec
     x,\tau)\right) d x^a dx^b \right],
\end{equation}
for a Friedman-Robertson-Walker Universe with scale factor $a(\tau)$, with small-amplitude synchronous-gauge perturbations $h_{ab}(\vec x,\tau)$.  The metric perturbation can be decomposed into scalar, vector, and transverse-traceless tensor components, each of which evolves, at linear order in the perturbation amplitude, independently.  We will deal with scalar perturbations (density perturbations) and tensor modes (gravitational waves).  Below we will need the conformal Hubble parameter ${\cal H}(\tau)\equiv\dot a/a$, where the dot denotes derivative with respect to conformal time, and also the scale factor $R(\tau)
\equiv (3/4) \rho_b(\tau)/\rho_\gamma(\tau)$ in units of $3/4$ of that at photon-baryon equality; $\rho_b(\tau)$ and $\rho_\gamma(\tau)$ are mean baryon and photon energy densities, respectively.

\subsection{Photon intensity matrix}
\label{sec:photonintensity}

At any given point $\vec x$ and time $\tau$, perturbations to the photon distribution function can be described in terms of a temperature fluctuation $T(\vec x,\hat n,\tau)$ for light coming from direction $\hat n$ and the Stokes parameters (measured with respect to two unit vectors $\hat \theta$ and $\hat\phi$ orthogonal to $\hat n$) $Q(\vec x,\hat n,\tau)$ and $U(\vec x,\hat n,\tau)$.

Following \cite{Weinberg:2008zzc} (see also \cite{Hu:1997hp}) we define an ``intensity matrix'' $J_{ab}(\vec x,
\hat n,\tau)$ to encode these perturbations to the photon distribution function.\footnote{Our $J_{ab}$ is 1/4 of that in \cite{Weinberg:2008zzc}, which is a {\it bona fide} intensity.}  The matrix is everywhere orthogonal to $\hat
n$ ($\hat n^a J_{ab}=0$) and symmetric
($J_{ab}=J_{ba})$ and can be written,
\begin{equation}
     J_{ab}(\vec x, \hat n,\tau) =\frac{1}{2 } T(\vec x,\hat n,\tau) \Gamma_{ab}(\hat n)
     + \frac{1}{\sqrt{2}} {\cal P}_{ab}(\vec x,\hat n,\tau),
\label{eqn:intensitymatrix}
\end{equation}
in terms of the symmetric trace-free polarization tensor ${\cal P}_{ab}(\vec x,\hat n,\tau)$ \citep{Kamionkowski:2015yta} and a tensor $\Gamma_{ab} = g_{ab} - \hat n_a \hat n_b$ that projects onto the plane of the sky.  For example, for $\hat n$ in the $\hat z$ direction, the intensity matrix takes the form,
\begin{equation}
     J_{ab}(\vec x,\hat n=\hat z,\tau) = \frac{1}{2} \left( \begin{array}{ccc} T(\vec x, \hat z, \tau)+ Q(\vec x, \hat z,\tau) & U(\vec x, \hat z, \tau)  & 0\\ U(\vec x, \hat z, \tau) & T(\vec x, \hat z, \tau)- Q(\vec x, \hat z,\tau)  & 0 \\ 0 & 0 & 0 \\ 
     \end{array} \right).
\label{eqn:Jmatrix}     
\end{equation}
The matrix in any other direction can be obtained by applying the appropriate rotation matrix (see, e.g., Section 7.4 in \cite{Weinberg:2008zzc}).  \comment{Defined this way, the fractional temperature perturbation is the trace $T(\vec x,\hat n,\tau)/\bar T=J^c_{\ c}(\vec x,\hat n,\tau)$, and the polarization given by the trace-free components.}

Given that $\Gamma^a{}_a=2$, the trace $J^c{}_c(\vec x,\hat n,\tau) = T(\vec x,\hat n,\tau)$.  The temperature monopole and dipole are
\begin{equation}
\Delta^T_0(\vec x,\tau) = \int \frac{d\hat n}{4\pi} J^c{}_c(\vec x,\hat n,\tau), \qquad {\rm and} \qquad
     {\cal I}_{a}(\vec x,t) = - \int \frac{ d\hat n}{4\pi}
     J^{c}{}_c(\vec x, \hat n,\tau) \hat n_a,
\label{eqn:photonflux}     
\end{equation}
which can be written $\Delta^T_0(\vec x,\tau)=(4\pi)^{-1/2} T_{00}(\vec x, \tau)$ and ${\cal I}_a(\vec x,\tau) = -(12\pi)^{-1/2}\sum_{m=-1}^1 T_{1m}(\vec x,\tau) \hat e^m_a$, in terms of temperature multipole moments (where $\hat e^m_a$ are spherical unit vectors).  We write the dipole in terms of a photon velocity potential  $\theta_\gamma(\vec x,\tau) = 3 \nabla^a {\cal I}_a$.

We define a photon-intensity tensor,
\begin{equation}
     {\cal T}_{ab}(\vec x,\tau) = \int \frac{ d\hat n}{4\pi}
     J_{ab}(\vec x, \hat n,\tau) = \frac13 \Delta^T_0(\vec x,\tau) g_{ab} + \frac12\sqrt{\frac23}
      \Pi_{ab}(\vec x,\tau),
\label{eqn:intensity}     
\end{equation}
to be the angle average of the photon distribution function.  It is written in terms of the monopole $\Delta^T_0(\vec x, \tau)$ and a trace-free component $\Pi_{ab}(\vec x,\tau)$.  This can be expressed as $\Pi_{ab}(\vec x,\tau) = -\Delta^T_{2,ab}(\vec x,\tau) + \sqrt{6} \Delta^E_{2,ab}(\vec x,\tau)$ in terms of temperature and E-mode-polarization quadrupole fields,
\begin{eqnarray}
     \Delta^T_{2,ab}(\vec x,\tau) &\equiv& \sqrt{\frac32}\int \frac{d\hat n}{4\pi} J^c{}_c(\vec x,\hat n,\tau) \hat n_a \hat n_b =  \sqrt{\frac{1}{20\pi}} \sum_{m=-2}^{m=2} T_{2m}(\vec x,\tau) \hat t^m_{ab}, \nonumber \\
     \Delta^E_{2,ab}(\vec x,\tau) &\equiv&  \frac{1}{\sqrt{2}}
\int \frac{d \hat n}{4\pi} {\cal P}_{ab}(\vec x,\hat n,\tau)  = \sqrt{\frac{1}{20\pi}} \sum_{m=-2}^{m=2} E_{2m}(\vec x,\tau) \hat t^m_{ab},
\label{eqn:quadrupoles}
\end{eqnarray}
where we have used
\begin{equation}
     \int d \hat n \,Y_{2m}(\hat n) \hat n_a \hat n_b = 2 \sqrt{\frac{2\pi}{15}} \hat t^m_{ab}, \quad
     \int d \hat n\, Y^E_{(2m)ab}(\hat n) =  \sqrt{\frac{8\pi}{5}} \hat t^m_{ab}, \quad  \int d \hat n\, Y^B_{(2m)ab}(\hat n) = 0,
\end{equation}
and $\hat t^m_{ab}$ are normalized rank-2 spherical tensors.  It will be useful below to keep in mind that $\Delta^T_0(\vec x,\tau)$ is (1/4 of) the photon-density field, a scalar function of $\vec x$; ${\cal I}_a(\vec x,\tau)$ is the photon peculiar-velocity field, a vector (or rank-1 tensor) field on $\vec x$; and the quadrupole fields ${\cal T}_{ab}(\vec x,\tau)$, $\Pi_{ab}(\vec x,\tau)$, $\Delta^T_{2,ab}(\vec x,\tau)$ and $\Delta^E_{2,ab}(\vec x,\tau)$ are rank-2 tensor fields.

\subsection{Boltzmann equation}

The Boltzmann equation for the photon distribution function can be written [cf.~Eq~(6.1.17) in \cite{Weinberg:2008zzc}],
\begin{equation}
     \left( \frac{\partial}{\partial \tau} - \hat n \cdot \vec
     \nabla \right) J_{ab}(\vec x, \hat n,\tau) = - \frac14 \Gamma_{ab}(\hat
     n) \hat n^c \hat n^d \dot h_{cd}(\vec x,\tau) -  \frac12 \dot\kappa \Gamma_{ab}(\hat n) \hat n_c v^c(\vec x,\tau)  
     -\dot \kappa J_{ab}(\vec
     x,\hat n,\tau) + \frac32 \dot\kappa \Gamma_{a}{}^c(\hat n)
     \Gamma_{b}{}^d(\hat n) {\cal T}_{cd}(\vec x,\tau) .
\label{eq:Boltzmann}    
\end{equation}
Here $\dot \kappa$ is the derivative of the Thomson-scattering optical
depth with respect to conformal time.  The left-hand side of Eq.~(\ref{eq:Boltzmann}) is the total time derivative of the intensity tensor along the trajectory of a light ray propagating in the $-\hat n$ direction.  It thus contains the total time derivative of the temperature fluctuation (from the trace) and of the polarization (from the trace-free part).

The first term on the right-hand side of Eq.~(\ref{eq:Boltzmann}) describes the effects of propagation through a metric perturbation $h_{ab}(\vec x,\tau)$.  Since $\Gamma_{ab}(\hat n)$ is a metric in the two-dimensional plane transverse to $\hat n$, it has no traceless component, and so this term contributes only to temperature fluctuations, not the polarization.  The second term accounts for Doppler shifts induced by scattering from moving gas.
The vector field $v_a(\vec x,\tau)$ is the baryon peculiar
velocity, which for scalar perturbations we can write  in terms of a scalar velocity potential $\theta_{\rm b}(\vec x,\tau) = -i \nabla_a v_a(\vec x,\tau)$.  This is then related to the photon-temperature dipole
through
\begin{equation}
     \left[ \frac{\partial}{\partial \tau} + {\cal H}(\tau) +
     \frac{\dot\kappa}{R(\tau)} \right] \theta_{\rm b}(\vec x,\tau) =
     \frac{\dot \kappa}{R(\tau)} \theta_\gamma(\vec
     x,\tau).
\label{eqn:peculiarvelocity}     
\end{equation}
The peculiar-velocity term in Eq.~(\ref{eq:Boltzmann}) is also proportional to $\Gamma_{ab}(\hat n)$ and thus also contributes only to the temperature fluctuation, and not the polarization.

The third term in Eq.~(\ref{eq:Boltzmann}) describes the (Thomson) scattering of photons out of the direction $\hat n$ and affects the polarization and temperature similarly.  The fourth term, proportional to the photon-intensity tensor ${\cal T}_{ab}(\vec x,\tau)$, describes the scattering of photons from other directions into the direction $\hat n$.  It is discussed  in \ref{app:scattering}.  We include cosmic birefringence in Section \ref{sec:cb}.

To proceed, we integrate Eq.~(\ref{eq:Boltzmann}) along the trajectory of the ray to obtain an expression,
\begin{eqnarray}
     J_{ab}(\vec x=0,\hat n,\tau) &=& \int_{\tau_i}^\tau\, d\tau'
     \, e^{-\kappa(\tau,\tau')} \left\{  - \frac14 \Gamma_{ab}(\hat
     n) \hat n^c \hat n^d \dot h_{cd}(\hat n(\tau-\tau'),\tau')
     + \frac32 \dot\kappa(\tau') \Gamma_{a}{}^c(\hat n)
     \Gamma_{b}{}^d(\hat n) {\cal T}_{cd}(\hat n(\tau-\tau'),\tau')
     \right. \nonumber \\
     & & \ \ \ \ \ \left.  -  \frac12
     \dot\kappa(\tau') \Gamma_{ab}(\hat n) \hat n^c v_c(\hat n
     (\tau-\tau'),\tau') \right\} + J_{ab}(\hat n
     (\tau-\tau_i),\hat n,\tau_i),
\label{eqn:los}     
\end{eqnarray}
for the intensity perturbation at the coordinate origin (which
we take as our observation point).  Here, $\kappa(\tau,\tau') = \int_{\tau'}^\tau \, d\tau_1\, \dot\kappa(\tau_1)$
is the optical-depth change between $\tau'$ and $\tau$. 
Of course, the integrand in Eq.~(\ref{eqn:los})
contains the intensity perturbation, and so Eq.~(\ref{eqn:los}) is strictly speaking an integral equation for $J_{ab}(\vec x,\hat n,\tau)$.

\section{Integral equations for photons: Scalar perturbations}
\label{sec:scalar}

\subsection{Decomposition of metric and intensity tensors}

The metric perturbation can be written
\begin{equation}
     h_{ab}(\vec x,\tau) = \frac13 h(\vec x,\tau) g_{ab} + \sqrt{\frac23} W^L_{ab} \alpha(\vec x,\tau),
     \label{eqn:metricperturbation} 
\end{equation}
in terms of its trace $h(\vec x,\tau)$ and longitudinal
components $\alpha(\vec x,\tau)$.\footnote{The factor $\sqrt{2/3}$ is included so that $\alpha(\vec x,\tau)$, and similarly other perturbations variables, agree with prior literature, where longitudindal tensors are defined through the unnormalized operator $\nabla^{-2}([\nabla_a \nabla_b-(g_{ab}/3)\nabla^2]$, rather than $W^L_{ab}$, which is normalized to $W^{L\, ab} W^L_{ab}=1$.}  \comment{The $\eta$ variable in \cite{Ma:1995ey} (or $\alpha$ in \cite{Seljak:1996is}) is related to ours through $\eta= (\alpha-h)/6$.  And their conformal-Newtonian variables $\phi$ and $\psi$ are related to ours by $\nabla^2 \psi = - \left(\ddot \alpha + \frac{\dot a}{a} \dot \alpha \right)/2$ and $\nabla^2\phi=  \nabla^2(\alpha-h)/6 + (\dot a/a) \dot\alpha/2$.}
The intensity tensor can likewise be written
\begin{equation}
     {\cal T}_{ab}(\vec x,\tau) = \frac13 \Delta^T_0(\vec x,\tau) g_{ab} + \frac12 \sqrt{\frac23} W^L_{ab} \Pi(\vec x,\tau),
\label{eqn:intensity}     
\end{equation}
in terms of a trace $\Delta^T_0(\vec x, \tau)$ and longitudinal
component $\Pi(\vec x,\tau)$.  The factor of $1/2$  above is chosen so that our $\Pi(\vec x)$ agrees with that in prior work.

\subsection{Equations for photon distribution}

We now consider a single scalar TAM wave $\Psi_{klm}(\vec x,\tau)$ with quantum numbers $klm$.  We write 
\begin{eqnarray}
    h(\vec x,\tau) &=&  h_{klm}(\tau) \Psi_{(klm)}(\vec x), \qquad \Delta^T_{0}(\vec x,\tau) = \Delta^T_{0,klm}(\tau) \Psi_{(klm)}(\vec x), \nonumber \\
    W^L_{ab} \alpha(\vec x,\tau)& =&  \alpha_{klm}(\tau) \Psi^L_{(klm)ab}(\vec x), \qquad  W^L_{ab} \Pi(\vec x,\tau)= \Pi_{klm}(\tau) \Psi^L_{(klm)ab}(\vec x),
\label{eqn:scalarTAMexpansions}
\end{eqnarray}
and now use Eq.~(\ref{eqn:los}) to determine the temperature/polarization pattern induced by this TAM wave.  We start with the trace $J_{a}{}^a(\vec x=0,\hat n,\tau)$,
the temperature fluctuation in direction $\hat n$, and
evaluate its spherical-harmonic coefficients.  For economy of notation, we temporarily omit factors of $4\pi i^l$ in the expansions from Eq.~(\ref{eqn:TAMexpansions}).  They will then be re-inserted into the expressions for CMB power spectra.

\subsubsection{Temperature fluctuation
}
Using Eq.~(\ref{eqn:scalarradial}), we find that for this $klm$ TAM wave, 
\begin{equation}
     \hat n^a \hat n^b \dot h_{ab}(\hat n, r,\tau) = \left[ \frac13
     \dot h_{klm}(\tau)
     j_l(kr) + \frac23 \dot \alpha_{klm}(\tau) R^{L,L}_l(kr) \right]
     Y_{lm}(\hat n).
\end{equation}
Likewise, for this TAM wave,
\begin{equation}
     \Gamma^{a}{}_c(\hat n) \Gamma_{ad}(\hat n) {\cal T}^{cd}(\hat n r,\tau) = 
     \Gamma_{cd}(\hat n) {\cal T}^{cd} = \left[ \frac23 \Delta^T_{0,klm}(\tau)
     j_l(kr) -\frac13 \Pi_{klm} R^{L,L}_l(kr) \right]
     Y_{lm}(\hat n).
\label{eqn:scatteringterm}     
\end{equation}
The last term in Eq.~(\ref{eqn:los}), the Doppler contribution, is evaluated using $\Gamma^a{}_a(\hat n) = 2$, $\theta_{{\rm b},klm}(\tau)=-ik v_{klm}(\tau)$, and $n^a \nabla_a =(\partial/\partial r)$.  Then for this $klm$ mode,
\begin{equation}
     \hat n^a v_a(\vec x,\tau) = v_{klm}(\tau) n^a \Psi^L_{(klm)a}(\vec x) = (i/k) v_{klm}(\tau) n^a \nabla_a \Psi{(klm)}(\vec x) = -[\theta_{{\rm b},klm}(\tau)/k] Y_{(lm)}(\hat n) j_l'(kr),
\label{eqn:dopplerprojection}
\end{equation}
where $j_l'(x)$ is the derivative with respect to the argument.

All the contributions to the integral in Eq.~(\ref{eqn:los}) are proportional to $Y_{(lm)}(\hat n)$, as they should be.  And given that $T(\hat n) = \sum_{lm} T_{lm} Y_{(lm)}(\hat n)$, we find that this $klm$ mode induces
a temperature spherical-harmonic coefficient,
\begin{eqnarray}
     T_{klm}(\vec x=0,\tau) &=&  \int_{\tau_i}^{\tau} \, d\tau' \,
     e^{-\kappa(\tau, \tau')} \left\{  \left[ -\frac16 \dot
     h_{klm}(\tau') + \dot\kappa \Delta^T_{0,klm}(\tau') \right]
     j_l( w) \right. \nonumber \\
       &  & \left. - \left[\frac13 \dot\alpha_{klm}(\tau') + \frac12
      \dot\kappa \Pi_{klm}(\tau') \right] R^{\rm
     L,L}_l(w) + \dot\kappa \theta_{{\rm b},klm}(\tau') j_l'(w)/k
     \right\} + T_{klm}(\vec x=0,\tau_i),
\label{eqn:Tfluctuation}
\end{eqnarray}
with $w=k(\tau-\tau')$.  The radial eigenfunctions in this approach are seen to be the projections of the appropriate TAM waves along and perpendicular to the line of sight.  It is gratifying to see that a given observable of quantum numbers $lm$ is determined by a three-dimensional perturbation that is an eigenstate of rotations with the same quantum numbers.

To be complete, we include on the right-hand side of Eq.~(\ref{eqn:Tfluctuation}) the initial value $T_{klm}(\vec x=0,\tau_i)$.  However, in subsequent integral equations, we will leave this initial value out, to de-clutter equations.  \cite{Kamionkowski:2021njk} discusses how this initial condition, as well as the initial conditions for the integral equation, can be specified in numerical work.

Now consider a TAM wave with $l=m=0$.  From Eq.~(\ref{eqn:photonflux}),
$\Delta^T_0(\vec x=0,\tau) = (4\pi)^{-1/2} T_{k00}(\vec x,\tau)$ for this $klm$ mode. However, we also have for an $l=m=0$ TAM wave $\Delta^T_0(\vec x=0,\tau)=\Delta^T_{0,k00}(\tau) \Psi_{(k00)}(\vec x=0)= (4\pi)^{-1/2} \Delta^T_{0,k00}$.  We thus infer that $\Delta^T_{0,k00} = T_{k00}(\tau)$.  Recall now that the time dependence of the perturbations of all TAM modes of the same $k$ are independent of $lm$.  If we suppress the $lm$ subscripts on the perturbation variables in Eq.~(\ref{eqn:Tfluctuation}), then we obtain for $l=0$ an integral representation for the time evolution of the photon-temperature monopole $\Delta^T_{0,k}(\tau)$ for any perturbation mode of wavenumber $k$.

Similarly, for $l=2$, the temperature quadrupole defined in Eq.~(\ref{eqn:quadrupoles}) is, at the origin, $\Delta^T_{2,ab}(\vec x=0,\tau)=\Delta^T_{2,k2m}(\tau) \Psi^L_{(k2m)ab}(\vec x=0)=(20\pi)^{-1/2} \Delta^T_{2,k2m} \hat t^m_{ab}$.  Comparing with the expression for $\Delta^T_{2,ab}(\vec x=0,\tau)$ in Eq.~(\ref{eqn:quadrupoles}), we infer that $\Delta^T_{2,k2m} =\, T_{k2m}(\vec x=0,\tau)$.  We can again suppress the $lm$ subscripts in Eq.~(\ref{eqn:Tfluctuation}) and set $l=2$ in Eq.~(\ref{eqn:Tfluctuation}) to obtain an integral for the time evolution of the temperature quadrupole $\Delta^T_{2,k}(\tau)$ for perturbations of wavenumber $k$.  It also follows, using $\Psi^L_{(klm)a}(\vec x=0) = i (12\pi)^{-1/2} \delta_{l1} \hat e^m_a$ that $\theta_{\gamma,k1m}(\tau) = 3 k T_{k1m}$ and thus, more generally, that we can write $\theta_{\gamma,k}=3 k \Delta_{1,k}(\tau)$.

The CMB temperature power spectrum $C_l^{TT}$ consists of expectation values of $|T_{lm}|^2$ and can be written as an integral $C_l^{TT} = (2/\pi) \int k^2\, dk\, |\Delta^T_{l,k}|^2$.\footnote{The prefactor $(2/\pi)$ comes from the product of the factor $(2\pi)^{-3}$ in the measure for $k$ and a factor $(4\pi)^2$ from the fact that the normalized eigenfunctions are $4\pi i^l \Psi(\vec x)$.}  Given Eq.~(\ref{eqn:Tfluctuation}), we thus infer that the temperature transfer functions are
\begin{equation}
     \Delta^T_{l,k}(\tau) = \int_{\tau_i}^{\tau} \, d\tau' \,
     e^{-\kappa(\tau, \tau')} \left\{  \left[ -\frac16 \dot
     h_{k}(\tau') + \dot\kappa \Delta^T_{0,k}(\tau') \right]
     j_l( w) - \left[\frac13 \dot\alpha_{k}(\tau') + \frac12
      \dot\kappa \Pi_{k}(\tau') \right] R^{
     L,L}_l(w) + \dot\kappa \theta_{{\rm b},k}(\tau') j_l'(w)/k
     \right\}.
\label{eqn:TTransferfunctions}
\end{equation}

\subsubsection{Polarization}

The polarization observed at the origin in direction $\hat n$ is 
\begin{equation}
{\cal P}_{ab}(\vec x=0,\hat n,\tau) = \sqrt{2} \left( J_{ab} - \frac12 \Gamma_{ab} J_{c}{}^c \right)(\vec x=0,\hat n,\tau).
\end{equation}
For a single $klm$ TAM wave, this is (again with no cosmic birefringence)
\begin{equation}
  {\cal P}_{ab}(\vec x=0,\hat n,\tau)_{klm} = \sqrt{2}\sqrt{\frac23} \int_{\tau_i}^\tau\, d\tau'\, e^{-\kappa(\tau,\tau')} \, \frac32 \dot\kappa\,
   \frac12 \Pi_{klm}(\tau')\left( \Gamma_a{}^c\ \Gamma_b{}^d-(1/2)\Gamma_{ab}\Gamma^{cd}  \right) \Gamma_e{}^c \Gamma_f{}^d\
          \Psi^L_{klm(cd)}(\hat n (\tau-\tau'),\tau').
\label{eqn:Peqn}
\end{equation}
This singles out the TE-spherical-harmonic term in the last relation in Eq~({\ref{eqn:scalarprojections}), from which we infer
that the E-mode-polarization spherical-harmonic amplitude is
\begin{equation}
     E_{klm}(\vec x=0,\tau) = -\frac{\sqrt{3}}{2} \int_{\tau_i}^\tau\,
     d\tau'\, e^{-\kappa(\tau,\tau')} \dot\kappa \Pi_{klm}(\tau') R^{L,TE}_l(w).
\label{eqn:Eklm}
\end{equation}
This indicates that, as is well known, the polarization is determined by Thomson scattering of a quadrupole.

As above, rotational invariance allows us to infer that the polarization transfer functions $\Delta^E_{l,k}(\tau)$ needed to obtain the polarization power spectra $C_l^{EE}= (2/\pi)\int\, k^2\, dk |\Delta^E_{l,k}|^2$ are
\begin{equation}
     \Delta^E_{l,k}(\tau) =  -\frac{\sqrt{3}}{2} \int_{\tau_i}^\tau\,
     d\tau'\, e^{-\kappa(\tau,\tau')} \dot\kappa \Pi_{k}(\tau') R^{L,TE}_l(w).
\label{eqn:Ptransferfunctions}
\end{equation}
For $l=2$, this provides an integral representation of the E-mode quadrupole $\Delta^E_{2,k}(\tau)$.

Using $\Pi_k(\tau)= -\Delta^T_{2,k}(\tau)+\sqrt{6} \Delta^E_{2,k}(\tau)$, Eqs.~(\ref{eqn:TTransferfunctions}) and (\ref{eqn:Ptransferfunctions}) provide a set of coupled integral equations for the time evolution of the photon monopole, dipole, and quadrupole that appear as sources for the Einstein equations for the metric-perturbation variables $h_k(\eta)$ and $\alpha_k(\eta)$.  They can be used in lieu of the usual Boltzmann hierarchy \citep{Weinberg:2006hh,Kamionkowski:2021njk}.  The $\Delta^E_{2,k}(\tau)$ equation can alternatively be replaced by an integral equation,
\begin{equation}
     \Pi_k(\tau) = -\Delta^T_{2,k}(\tau) + 9 \int_{\tau_i}^\tau\, d\tau' \, e^{-\kappa(\tau,\tau')} \dot \kappa \Pi_k(\tau') \frac{j_2(w)}{w^2},
\label{eqn:simplerPieqn}     
\end{equation}
for the scattering source $\Pi_k(\tau)$.
As shown in \ref{app:boltzmann}, differentiation of Eqs.~(\ref{eqn:TTransferfunctions}) and (\ref{eqn:Ptransferfunctions}) equations with respect to the conformal time $\tau$ recovers the usual Boltzmann equations.

\subsection{Auxiliary equations}

Our primary aim was the TAM derivation of the integral equations for photon intensity and polarization, but we briefly describe the remaining equations needed to evolve perturbations.  In the traditional approach to cosmological perturbations, an infinite tower of Boltzmann equations for the photon temperature and polarization moments are solved to obtain the temperature monopole $\Delta^T_{k0}$, dipole $\Delta^T_{k1}$ (or photon velocity $\theta_\gamma$), and temperature quadrupole $\Delta^T_{2,k}$, which then communicate to the rest of the perturbations through the Einstein equations and the equation for photon-baryon drag.  In the integral-equation approach, they are obtained by solving Eqs.~(\ref{eqn:Tfluctuation}) and (\ref{eqn:simplerPieqn}) simultaneously for $l=0$, $l=1$, and $l=2$.\footnote{Strictly speaking, once $\Delta^T_{0,k}(\tau)$ is obtained from the integral equation, $\theta_{\gamma k}(\tau)= 3 k \Delta^T_{1,k}(\tau)$ and $\Delta^T_{2,k}(\tau)$ can alternatively be obtained from the coupling equations, $\dot \Delta^T_{0} = -\theta_\gamma/3-\dot h/6$ and $\dot\theta_{\gamma k} = \dot\kappa (\theta_{{\rm b}k}-\theta_{\gamma_k}) + k^2 (\Delta^T_{k0} - 2\Delta^T_{k2})$, although the integral equations may in some cases be numerically more stable \citep{Kamionkowski:2021njk,Ji:2022iji,Lee:2025zym}.}

The metric perturbations that appear in Eq.~(\ref{eqn:Tfluctuation}) are obtained from the Einstein equations,
\begin{equation}
     \ddot h +  \frac{\dot a}{a} \dot h = - 8 \pi G a^2 \left[ \delta \rho_{\rm tot} + 3 \delta p_{\rm tot} \right],\qquad \mathrm{and} \qquad  \frac13 (\dot h - \dot \alpha) = 8 \pi G a^2 (\bar \rho_{\rm tot}+ \bar p_{\rm tot})\theta_{\rm tot},
\label{eqn:firsteinstein}     
\end{equation}
in terms of the total density perturbation $\delta\rho_{\rm tot}$, pressure perturbation $\delta p_{\rm tot}$, and momentum flux $ (\bar \rho_{\rm tot}+ \bar p_{\rm tot})\theta_{\rm tot}$.  

The equation of motion for the fractional cold-dark-matter perturbation $\delta_{\rm c}(\vec x,\tau)$ is $\dot\delta_{\rm c}  = -\dot h/2$ (assuming dark matter is collisionless).
Synchronous-gauge coordinates are those that follow freely-falling observers, and so the dark matter has no peculiar velocity in this gauge.  The fractional baryon-density perturbation $\delta_{\rm b}$ satisfies $\dot\delta_{\rm b} + \theta_{\rm b} = -\dot h/2$.  The baryon velocity $\theta_{\rm b}$ is coupled to the photons through Eq.~(\ref{eqn:peculiarvelocity}).

Neutrinos interact only gravitationally, and only the monopole, dipole, and quadrupole (density, peculiar velocity, and anisotropic stress) of the neutrino distribution function communicate with the gravitational fields.  For massless neutrinos, the integral equations for these three quantities are obtained from the analogous equations for photons by (1) taking the collisionless limit $\dot\kappa\to0$ and (2) ignoring the polarization.  The case of massive neutrinos is detailed in \cite{Ji:2022iji}.

\section{Adding cosmic birefringence}
\label{sec:cb}

The Boltzmann equation for the photon distribution function including cosmic birefringence is
\begin{eqnarray}
     \left( \frac{\partial}{\partial \tau} - \hat n \cdot \vec
     \nabla \right) J_{ab}(\vec x, \hat n,\tau) &=& - \frac14 \Gamma_{ab}(\hat
     n) \hat n^c \hat n^d \dot h_{cd}(\vec x,\tau) -  \frac12 \dot\kappa \Gamma_{ab}(\hat n) \hat n_c v^c(\vec x,\tau)  \nonumber \\
     & & 
     -\dot \kappa J_{ab}(\vec
     x,\hat n,\tau)
     + \frac32 \dot\kappa \Gamma_{a}{}^c(\hat n)
     \Gamma_{b}{}^d(\hat n) {\cal T}_{cd}(\vec x,\tau) 
     + \dot \beta (\epsilon_{a}{}^{dc} \hat n_d J_{cb} +\epsilon_{b}{}^{dc} \hat n_d J_{ac}).
\label{eq:BoltzmannCB}    
\end{eqnarray}
Here  $\dot \beta$ is the derivative of the cosmic-birefringence rotation angle $\beta$, and $\epsilon_{abc}$ is the antisymmetric tensor.   
 
The last term describes the effects of cosmic birefringence, a rotation of the observed polarization orientation in the clockwise direction (for $\dot\beta>0$).  Such a rotation arises, for example, if there is a term of the form $-g\phi F_{\mu\nu}\widetilde F^{\mu\nu}/4$, where $\phi(\tau)$ is a uniform time-evolving scalar field, added to the Lagrangian for the electromagnetic field $F_{\mu\nu}$.  In this case $\dot\beta = g \dot \phi/2$.

The cosmic-birefringence (last) term in Eq.~(\ref{eq:BoltzmannCB}) augments the line-of-sight solution for the intensity matrix, Eq.~(\ref{eqn:los}), to
\begin{eqnarray}
    J_{ab}(\vec x = 0,\hat n,\tau) &=& \int_{\tau_i}^{\tau} d\tau'\, e^{-\kappa(\tau,\tau')} \left[ -\frac14 \Gamma_{ab}(\hat n)\hat n^e\hat n^f \dot h_{ef} + \frac32  \Theta_a{}^c(
    \beta(\tau,\tau');\hat n) \Theta_b{}^d(\beta(\tau,\tau');\hat n)\dot\kappa(\tau')\Gamma_{ce}\Gamma_{df}\mathcal T^{ef} \right. \nonumber \\
    &+& \left. \frac12 \dot\kappa(\tau')\Gamma_{ab}\hat n^e v_e \right] + \Theta_a{}^c(\beta(\tau,\tau_i);\hat n)\Theta_b{}^d(\beta(\tau,\tau_i);\hat n) J_{cd}(\hat n(\tau-\tau_i),\hat n,\tau_i),
\label{eqn:cblos}    
\end{eqnarray}
where
\begin{equation}
    \Theta_a{}^c(\beta;\hat n) = \hat n_a \hat n^c + \cos\beta\;\Gamma_a{}^c(\hat n) + \sin\beta\;\epsilon_a{}^{ec}\hat n_e,
\label{eqn:rotationoperator}    
\end{equation}
and $\beta(\tau,\tau')=\int_{\tau'}^\tau\, d\tau_1 \dot\beta(\tau_1)$.  The projection operator $\Theta_a{}^c$ rotates the traceless part of the tensor it acts on about the direction $-\hat n$ of propagation by $\beta$, in a left-handed sense.

The affected scattering term appears in the derivation of the temperature transfer function [Eq.~(\ref{eqn:los})] through the scattering term in Eq.~(\ref{eqn:scatteringterm}).  As seen in that equation, that scattering term depends on the trace of the components of the intensity tensor ${\cal T}_{ab}$ transverse to $\hat n$.  However, $\Theta_{ab}$ simply rotates the components of ${\cal T}_{ab}$ transverse to $\hat n$ and thus leaves the trace unaffected.  Eq.~(\ref{eqn:scatteringterm}) for the temperature transfer function is thus unaffected by cosmic birefringence.

Consider now Eq.~(\ref{eqn:Peqn}) for the polarization.  The factor $\Gamma_{ac}(\hat n)$ is replaced by $\Theta_{a}{}^{e}\Gamma_{ec}$ and similarly for $\Gamma_{bd}$.  This then replaces the E-mode harmonic $Y^{\rm E}_{(lm)ab}(\hat n)$ that would otherwise emerge by $\cos2\beta(\tau,\tau') Y^{\rm E}(\hat n) + \sin 2\beta(\tau,\tau') Y^{\rm B}(\hat n)$, thus introducing a non-zero B-mode polarization, as expected.  Eq.~(\ref{eqn:Eklm}) is thus augmented to
\begin{equation}
     \Delta^E_{kl}(\tau) + i \Delta^B_{kl}(\tau) = - \frac{\sqrt{3}}{2} \int_{\tau_i}^\tau\,
     d\tau'\, e^{-\kappa(\tau,\tau')} \dot\kappa \Pi_k(\tau') e^{2i\beta(\tau,\tau')} R^{L,TE}_l(w) +
     e^{2i\beta(\tau,\tau_i)} E_{klm}(\tau_i),
\label{eqn:EBklm}
\end{equation}
in agreement with earlier work \citep{Liu:2006uh,Murai:2022zur,Finelli:2008jv,Gubitosi:2014cua}.  Note that we have re-introduced the initial value explicitly to emphasize that it is rotated, but we leave it out of future expressions.

As before, the source $\Pi_k(\tau)=-\Delta^T_{2,k}(\tau)+\sqrt{6} \Delta^E_{2,k}(\tau)$, and so Eq.~(\ref{eqn:simplerPieqn}) becomes
\begin{equation}
     \Pi_k(\tau) = -\Delta^T_{2,k}(\tau) + 9 \int_{\tau_i}^\tau\, d\tau' \, e^{-\kappa(\tau,\tau')} \cos\left[2\beta(\tau,\tau')\right]\dot \kappa \Pi_k(\tau') \frac{j_2(w)}{w^2}.
\label{eqn:PieqnwithCB}     
\end{equation}
Although Eq.~(\ref{eqn:TTransferfunctions}) for the temperature transfer function is unaffected by cosmic birefringence, the source function $\Pi_k(\tau)$ is indeed affected.  As a result, cosmic birefringence will have some (probably small) effect on the CMB temperature which, as far as we know, has not been discussed in the prior literature.   We leave numerical investigation of this effect to future work.

\section{Integral equations for photons:  Tensor modes}
\label{sec:tensors}

We will now obtain expressions for the temperature and E- and B-mode polarization spherical-harmonic coefficients generated by ``tensor modes,'' transverse-traceless metric perturbations (gravitational waves). The derivation parallels that for scalar perturbations.

We expand the tensor metric perturbation and intensity tensor as
\begin{equation}
    h_{ab}(\vec x,\tau) = \sum_{\alpha={\rm TE,TB}}   h^\alpha_{klm}(\tau) \Psi^\alpha_{(klm)ab}(\vec x), \qquad       {\cal T}_{ab}(\vec x,\tau) = -\frac16 \sum_{\alpha={\rm TE,TB}}   \Pi^\alpha_{klm}(\tau) \Psi^\alpha_{(klm)ab}(\vec x),
\label{eqn:tensorexpansions}    
\end{equation}
where $\alpha$ sums over TE and TB tensor TAM.  The intensity tensor ${\cal T}_{ab}(\vec x,\tau)$ has no traceless component with tensor modes, and there is no velocity field induced by tensor perturbations and thus no contribution to the intensity matrix from the last term in Eq.~(\ref{eqn:los}).  We have chosen the factor $-1/6$ in the relation between ${\cal T}_{ab}(\vec x,\tau)$ and $\Pi^\alpha_{klm}(\tau)$ so that, as we will see below, our $\Pi^E(\tau)$ recovers what is usually called $\Psi(\tau)$ in prior work.  Since we still have ${\cal T}_{ab}(\vec x,\tau))  = - \Delta^T_{2,ab}(\vec x,\tau)/\sqrt{6} + \Delta^E_{2,ab}(\vec x,\tau)$, in terms of the quadrupoles in Eq.~(\ref{eqn:quadrupoles}), we now have $\Pi_{klm}^E(\vec x,\tau) =\sqrt{6} \left[ \Delta^T_{2,klm}(\vec x,\tau) -\sqrt{6} \Delta^E_{2,klm}(\vec x,\tau) \right]$.  

The last subtlety is that $\Psi^{TE}_{(k2m)ab}(\vec x=0)= \sqrt{2} \Psi^{L}_{(k2m)ab}(\vec x=0)$ [cf.~the first terms in Eqs.~(74) and (78) in \cite{Dai:2012bc}].  Thus, for tensor modes, the temperature and polarization-quadrupole amplitudes $\Delta^T_{2,klm}(\tau)$ and $\Delta^E_{2,klm}(\tau)$ are smaller by a factor of $\sqrt{2}$ than the temperature/polarization spherical-harmonic coefficients $T_{klm}(\vec x=0,\tau)$ and $E_{klm}(\vec x=0,\tau)$ that we are about to calculate.

\subsection{E modes}

We start with the temperature and polarization spherical-harmonic coefficients generated by an E mode of quantum numbers $klm$.  Consider first the temperature fluctuation obtained from the trace of Eq.~(\ref{eqn:los}).  Using Eq.~(\ref{eqn:tensorprojections}), $\Gamma^{ab} Y_{(lm)ab}^E(\hat n)=0$, and $\Gamma^a{}_a=2$, we find that a given $klm$ E-mode TAM wave induces a temperature fluctuation,
\begin{equation}
    T_{klm}(\vec x=0,\tau) =  \int_{\tau_i}^\tau\, d\tau'\, e^{-\kappa(\tau,\tau')} \left[\frac{1}{\sqrt{6}}\dot h_{klm}^E(\tau') - \frac{1}{2\sqrt{6}} \dot\kappa(\tau') \Pi^E_{klm}(\tau') \right] R_l^{L,TE}(w),
\label{eqn:EmodeT}    
\end{equation}
where again $w=k(\tau-\tau')$. After integrating over angles, the trace-free part of Eq.~(\ref{eqn:los}) gives, using Eq.~(\ref{eqn:tensorprojections}),
\begin{equation}
    E_{klm}(\vec x=0,\tau) = \frac{1}{2\sqrt{2}} \int_{\tau_i}^\tau\, d\tau'\,  \dot \kappa(\tau') e^{-\kappa(\tau,\tau')} \Pi^E_{klm}(\tau') R_l^{E,E}(w),
\label{eqn:EmodeE}    
\end{equation}
to the polarization spherical-harmonic coefficient from this $lm$ mode.

We now write temperature/polarization transfer functions,
\begin{equation}
    \Delta^T_{l,k}(\tau) = \frac{1}{\sqrt{6}} \int_{\tau_i}^\tau\, d\tau'\, e^{-\kappa(\tau,\tau')} \left[ \dot h_{k}^E(\tau') - \frac12\dot\kappa(\tau') \Pi^E_{k}(\tau') \right] R_l^{L,TE}(w),
\label{eqn:TensorDeltaT}    
\end{equation}
\begin{equation}
    \Delta^E_{l,k}(\tau) = \frac{1}{2\sqrt{2}} \int_{\tau_i}^\tau\, d\tau'\,  \dot \kappa(\tau') e^{-\kappa(\tau,\tau')} \Pi^E_{k}(\tau') R_l^{E,E}(w),
\label{eqn:TensorDeltaE}
\end{equation}
so that the CMB power spectra are $C_l^{XX}=(2/\pi)\int\, k^2\, dk\, |\Delta^{X}_{l,k}|^2$, for $X=T,E$.\footnote{Our expressions for $C_l^{TT}$ and $C_l^{EE}$ appear to be smaller by factors of 2 and 6, respectively, than those in \cite{Weinberg:2008zzc} and \cite{Hu:1997hp}, respectively.  This apparent discrepancy stems from the fact that their normal modes involve unit tensors normalized to 2 and 6, respectively, while our normal modes are TAM waves normalized to unity.}

With these definitions of $\Delta^T_{l,k}$ and $\Delta^E_{l,k}$, we now set $\Pi^E_k(\tau) = \sqrt{3}(\Delta^T_2-\sqrt{6} \Delta^E_2)$, taking into account the factor of $\sqrt{6}$ between $\Pi_{klm}$ and $(\Delta^T_{2,klm}-\sqrt{6} \Delta^E_{2,klm})$ and the factor of $\sqrt{2}$ between those quadrupoles and the temperature/polarization spherical-harmonic coefficients in Eqs.~(\ref{eqn:TensorDeltaT}) and (\ref{eqn:TensorDeltaE}).   We then obtain an integral equation\footnote{Our result agrees with Eq.~(29) in \protect\cite{Weinberg:2006hh} by identifying ${\cal D}_q$ with our $h^E_k$ and $\Psi$ there with our $\Pi^E$ [see Eqs.~(15) and (20) there and our Eq.~(\protect\ref{eqn:tensorexpansions})]. We also need to identify $j_0(w) - 2 j_1(w)/w +  2j_2(w)/w^2 = j_2'(w)/w + ( 5/w^2-1) j_2(w) = - R_2^{L,TE}(w)/(3\sqrt{2})- R_2^{E,E}(w)$.  A similar integral equation also appeared in \protect\cite{Baskaran:2006qs}.} 
\begin{equation}
     \Pi^E_k(\tau) =   \frac{1}{\sqrt{2}} \int_{\tau_i}^\tau\, d\tau'\, e^{-\kappa(\tau,\tau')} \, \left[ \dot h^E_k(\tau') R_2^{L,TE}(w)-\frac12 \dot\kappa \Pi^E_k(\tau')\left( R_2^{L,TE}(w) + 3\sqrt{2} R_2^{E,E}(w) \right) \right].
\label{eqn:tensorTtransfer}     
\end{equation}
We can check the result for $T_{klm}$ with the 
Einstein equation,
\begin{equation}
    \ddot h_{ab}(\vec x,\tau) + 2\frac{\dot a}{a} \dot h_{ab}(\vec x,\tau) -\nabla^2 h_{ab}(\vec x,\tau) = 16 \pi G a^2 \pi_{ab}(\vec x,\tau),
\end{equation}
where $\pi_{ab}(\vec x,\tau)$ is the transverse-traceless part of the anisotropic stress.  The contribution of photons to the right-hand side is
\begin{equation}
   16\pi G\, \pi_{ab}(\vec x,\tau)= 64 \pi G \bar\rho_\gamma(\tau) \int \frac{d\hat n}{4\pi} \, J^c{}_c(\vec x,\hat n,\tau) \hat n_a \hat n_b = 24\, f_\gamma (\dot a/a)^2\sqrt{\frac23} \Delta^T_{2,ab}(\vec x,\tau),
\end{equation}
where we have used Eq.~(\ref{eqn:quadrupoles}) and $f_\gamma$ is the fraction of the total density contributed by photons.
We now consider one TAM wave of quantum number $klm$ and note that the coefficient for the TAM-wave expansion of the temperature-quadrupoole field $\Delta^T_{2,ab}(\vec x)$ is
\begin{equation}
    \Delta^T_{2,klm}(\vec x=0,\tau) = \frac{1}{\sqrt{2}} \int_{\tau_i}^\tau\, d\tau'\, e^{-\kappa(\tau,\tau')} \left[\frac{1}{\sqrt{6}}\dot h_{klm}^E(\tau') - \frac{1}{2\sqrt{6}} \dot\kappa(\tau') \Pi^E_{klm}(\tau') \right] R_2^{L,TE}(w).
\label{eqn:Delta2T}    
\end{equation}
Using $R^{L,TE}_2(w) = -3\sqrt{2} j_2(w)/w^2$, and assuming that only photons contribute to the anisotropic stress, we arrive at an integro-differential equation,
\begin{equation}
    \ddot h_{klm}^E(\tau) + 2\frac{\dot a}{a} \dot h^E_{klm}(\tau)+ k^2 h^E_{klm}(\tau) = 
    -24 f_\gamma \left(\frac{\dot a}{a}\right)^2\int_{\tau_i}^\tau\, d\tau'\, e^{-\kappa(\tau,\tau')} \left[\dot h^E_{klm}(\tau) \kappa(\tau') - \frac12 \dot\kappa(\tau') \Pi^E_{klm}(\tau') \right] \frac{j_2(w)}{w^2}.
\end{equation}
This agrees with the equation for the neutrino anisotropic stress \citep{Weinberg:2003ur} in the limit that $\dot\kappa\to0$.

\subsection{B modes}

We now consider the spherical-harmonic coefficients generated by a B mode. The polarization at the origin induced by a B-mode TAM wave of quantum numbers $klm$ is
\begin{equation}
    B_{klm}(\vec x=0,\tau) = \frac{1}{2\sqrt{2}} \int_{\tau_i}^\tau\, d\tau'\,  \dot \kappa(\tau') e^{-\kappa(\tau,\tau')} \Pi^B_{klm}(\tau') R_l^{B,B}(w),
\label{eqn:BmodeB}    
\end{equation}
implying, following prior reasoning, that the CMB B-mode power spectrum will be $C_l^{BB} = (2/\pi) \int\, k^2 \, dk\, \left| \Delta^B_{l,k} \right|^2$, with
\begin{equation}
    \Delta^B_{l,k}(\vec x=0,\tau) = \frac{1}{2\sqrt{2}} \int_{\tau_i}^\tau\, d\tau'\,  \dot \kappa(\tau') e^{-\kappa(\tau,\tau')} \Pi^B_k(\tau') R_l^{B,B}(w).
\label{eqn:DeltaB}    
\end{equation}
But consider now the source $\Pi^B_{k}(\tau)$.  There is no temperature fluctuation at $\vec x=0$ and no E-mode polarization with this B-mode TAM wave.  That does not, however, imply that $\Pi^B_{klm}=0$ in Eq.~(\ref{eqn:BmodeB}), as the TB tensor TAM wave for $l=2$ also vanishes at the origin, $\Psi^{TB}_{(k2m)ab}(\vec x=0)=0$.  Although $\Pi_{ab}(\vec x=0)=\Delta^T_{2,ab}(\vec x=0)=\Delta^T_{2,ab}(\vec x=0)=0$, these quadrupoles do not vanish everywhere.  Since the TE/TB TAM waves are not eigenstates of spatial translations, the TE/TB separation is not preserved if we choose a different origin \citep{Kamionkowski:2014faa}, and there will, in fact be temperature fluctuations and E-mode polarization fluctuations {\it away from the origin} with a TB TAM wave.  

While it would in principle be straightforward to evaluate these quadrupoles at some $\vec x\neq 0$, the calculation becomes quite complicated.  To proceed, we may instead expand our metric perturbation $h_{ab}(\vec x)$ and intensity tensor ${\cal T}_{ab}(\vec x)$ in terms of spin-2 TAM waves $\Psi^{\pm2}(\vec x) \equiv \left[ \Psi^{TE}_{(klm)ab}(\vec x) \pm i\Psi^{TB}_{(klm)ab}(\vec x) \right]/\sqrt{2}$ with expansion coefficients $h^{\pm2}(\vec \tau)$ and $\Pi^{\pm2}(\vec \tau)$.  We would then obtain expressions for $T_{klm}(\vec x=0,\tau)$ and $E_{klm}(\vec x=0,\tau)$ like those in Eqs.~(\ref{eqn:EmodeT}) and (\ref{eqn:EmodeE}), respectively, but with $h^E \to h^{\pm2}$ and $\Pi^E \to \Pi^{\pm2}$, and then an equation for $B_{klm}(\vec x=0,\tau)$ like Eq.~(\ref{eqn:BmodeB}), but with $\Pi^B \to \Pi^{\pm2}$.  We can then obtain an equation for $\Pi^B$ by obtained the TE/TB separation from the appropriate linear combination of the $\pm2$ modes and thus find that the scattering source for a B-mode TAM wave satisfies exactly the same equation, Eq.~(\ref{eqn:tensorTtransfer}), as the E-mode scattering source, but with $h^E \to h^B$ and $\Pi^E \to \Pi^B$.

\section{Conclusion}
\label{sec:conclusion}

We have re-derived the equations for the linear evolution of cosmological perturbations, and in particular the CMB temperature and polarization transfer functions, using the total-angular-momentum formalism of \cite{Dai:2012bc}.  In place of the usual decomposition into plane waves that are projected onto the spherical sky only at the end of the calculation, we work from the outset with TAM waves, which incorporate the rotational symmetry of the sky from the start.  As a consequence, a scalar TAM wave of quantum numbers $klm$ sources temperature and polarization only in the spherical-harmonic coefficients of the same $lm$, and the transfer functions emerge directly as line-of-sight integral equations rather than through the truncation and integration of a Boltzmann hierarchy.

The central results are the integral equations, Eqs.~(\ref{eqn:Tfluctuation}) and (\ref{eqn:simplerPieqn}), for the temperature transfer function and the radiation quadrupole $\Pi_k$ induced by scalar perturbations, together with the polarization transfer function, Eq.~(\ref{eqn:Eklm}), and the tensor-mode results of Section~\ref{sec:tensors}.  The TAM derivation makes the structure of these equations transparent.  The three source terms in the temperature transfer function are the line-of-sight projections of the scalar, longitudinal-vector, and longitudinal-tensor TAM waves, carrying the radial weights $j_l$, $j_l'$, and $R^{L,L}_l$, respectively, while polarization arises only from the Thomson scattering of the quadrupole and thus enters solely through the $E$-mode radial function $R^{L,TE}_l$.  These equations were stated without derivation in \cite{Kamionkowski:2021njk} and \cite{Ji:2022iji}; we have supplied that derivation here.  As a check, the Boltzmann hierarchy is recovered by differentiating the integral equations.

We have also incorporated cosmic birefringence.  For a uniform rotation, the line-of-sight solution acquires a rotation operator, Eq.~(\ref{eqn:rotationoperator}), that acts only on the polarization, rotating the emitted $E$ mode into a combination of $E$ and $B$; the result, Eq.~(\ref{eqn:EBklm}), reproduces earlier work.  Because the angular average of the $B$-mode tensor spherical harmonic vanishes, the induced $B$ mode does not feed back into the Thomson source, and the quadrupole equation, Eq.~(\ref{eqn:simplerPieqn}), is unchanged.  In \ref{app:aniso} we show that an anisotropic rotation angle is treated with similar economy in the TAM approach: a rotation field of total angular momentum $LM$ couples the polarization of a source multipole $l$ to the neighboring observed multipoles $l'$ in the band $|l-L|\le l'\le l+L$, through a single Gaunt coefficient obtained from the tensor-harmonic overlap integrals of \cite{Dai:2012ma}.

While our emphasis here has been on physical insight and on providing derivations absent from the literature, the integral-equation formulation also has practical value: it underlies the {\tt CLASSIER} \citep{Ji:2022iji,Lee:2025zym,Lee:2025vgv} and {\tt CLASSIER-DDM} \citep{Bencke:2026uws,Nanoominprep} codes, which replace the most costly Boltzmann hierarchies with more efficient integral-equation solutions.  The TAM treatment of cosmic birefringence developed here may prove similarly useful, both as a cross-check on existing treatments and as a starting point for the efficient computation of birefringence signatures, a subject of particular interest given the recent suggestion \citep{Namikawa:2025doa} that cosmic birefringence may relax neutrino-mass tensions with current data. 

\section*{Acknowledgments}
MK was supported by NSF Grant No.\ 2412361, NASA ATP Grant No.\ 80NSSC24K1226, and the Templeton Foundation Grant No.\ 62840.

\appendix

\section{Projections of TAM Waves}
\label{app:projections}

Eq.~(94) in \cite{Dai:2012bc} provides the tensor TAM waves in terms of radial eigenfunctions and projections onto tensor spherical harmonics.
From these we obtain projections onto the radial direction and transverse directions that will be needed in our calculation.  For scalar TAM waves, these are
\begin{align}
     \hat n^a \Psi^L_{(klm)a}(r\hat n)
     &= i\, j_l'(kr)\, Y_{(lm)}(\hat n), \qquad  \hat n^a\hat n^b \Psi^L_{(klm)ab}(r\hat n)
     = \sqrt{\frac23}\, R^{L,L}_l(kr)\, Y_{(lm)}(\hat n),
\nonumber\\
    & \Gamma_a{}^c(\hat n)\,\Gamma_b{}^d(\hat n)
     \Psi^L_{(klm)cd}(r\hat n)
     = - R^{L,TE}_l(kr)\, Y^{E}_{(lm)ab}(\hat n) -\frac{1}{\sqrt{6}} R_l^{L,L}(kr) \Gamma_{ab} Y_{(lm)}(\hat n),
\label{eqn:scalarprojections}
\end{align}
where $\Gamma_{ab}(\hat n) = g_{ab}-\hat n_a\hat n_b$, and the radial eigenfunctions are
\begin{equation}
     R^{L,L}_l(x) = \frac12\left[j_l(x) + 3 j_l''(x)\right],
     \qquad
     R^{L,TE}_l(x) = -\frac{\sqrt3}{2}
     \sqrt{\frac{(l+2)!}{(l-2)!}}\, \frac{j_l(x)}{x^2}. 
\label{eqn:scalarradial}
\end{equation}
For tensor metric perturbations, we will need the projections,
\begin{align}
     \hat n^a\hat n^b\, \Psi^{TE}_{(klm)ab}(r\hat n)
     &= -\sqrt{\frac23} R_l^{L,TE}(kr)
     Y_{(lm)}(\hat n), 
     \qquad
     \hat n^a\hat n^b\, \Psi^{TB}_{(klm)ab}(r\hat n) = 0, \nonumber \\
     \Gamma_a{}^c(\hat n)\Gamma_b{}^d(\hat n) \Psi^{TE}_{(klm)cd}(r\hat n)  
     &= R^{E,E}_l(kr) Y^E_{(lm)ab}(\hat n)  + \sqrt{\frac16} R_l^{L,TE}(kr)\Gamma_{ab}(\hat n) Y_{(lm)}(\hat n), 
    \nonumber \\
     \Gamma_a{}^c(\hat n)\Gamma_b{}^d(\hat n) \Psi^{TB}_{(klm)cd}(r\hat n)
     &= R^{B,B}_l(kr) Y^B_{(lm)ab}(\hat n), 
\label{eqn:tensorprojections}
\end{align}
with\footnote{We have changed the sign of $R^{E,E}_l(x)$  and also the sign of the $R^{L,TE}_l(x)$ term in Eq.~(\ref{eqn:tensorprojections}) since \cite{Dai:2012bc} use a different sign convention for $Y^E_{(lm)ab}(\hat n)$ than others.}
\begin{equation}
     R^{E,E}_l(x) = \frac12\left[-j_l(x) + j_l''(x)
     + 4\frac{j_l'(x)}{x} + 2\frac{j_l(x)}{x^2}\right],
     \qquad
     R^{B,B}_l(x) = -i\left[j_l'(x) + 2\frac{j_l(x)}{x}\right]. 
\label{eqn:tensorradial}
\end{equation}

\section{The scattering term}
\label{app:scattering}

Thomson scattering removes photons from the beam, as described by the $\dot\kappa J_{ab}$ term in Eq.~(\ref{eqn:los}).  Scattering of photons {\it into} the beam is described by the $(3/2) \dot \kappa \Gamma_{a}{}^c(\hat n) \Gamma_{b}{}^d(\hat n) {\cal T}_{cd}$ term.  This term can be understood as follows (see \cite{Weinberg:2008zzc} for more details):  The differential cross section for a photon with initial polarization vector $\hat \epsilon_i$ to scatter to a photon with polarization vector $\hat \epsilon_f$ is proportional to $(\hat\epsilon_f \cdot \hat \epsilon_i)^2$.  The intensity matrix $J_{ab}$ is proportional to $\sum_m p_m \hat e_a^m \hat e_b^m$, where $m$ sums over the two orthogonal polarizations (both orthogonal to the photon direction $\hat p$), and $p_m$ is the probability to find the photon in that polarization.  Since ${\cal T}_{cd}$ is an integral of $J_{cd}$ over all initial-state photon directions, the scattering term integrates the projection $\Gamma_{a}{}^c(\hat n) \Gamma_{b}{}^d(\hat n) J_{cd}$ over all initial-state directions $\hat p_i$.  The contraction of $\Gamma_a{}^c(\hat n) J_{cd}$ provides one factor of $\hat \epsilon_f \cdot \hat \epsilon_i$ and the contraction $\Gamma_b{}^d(\hat n) J_{cd}$ provides the other, both projected onto the components of the final-state intensity matrix for photons moving in direction $-\hat n$.  The factor of $3/2$ is a normalization that arises after integrating over angles.

\section{Boltzmann equations}
\label{app:boltzmann}

\subsection{Scalar equations}

\subsubsection{Temperature equations}

We start with Eq.~(\ref{eqn:TTransferfunctions})
and use $j_l'(w)=[ l j_{l-1}(w)-(l+1)j_{l+1}(w)]/(2l+1)$, $j_l(0)=\delta_{l0}$, $j_l'(0)=(1/3)\delta_{l1}$, and $R_l^{L,L}(0)=(1/5)\delta_{l2}$ to obtain (suppressing all subscripts $k$)
\begin{equation}
     \dot \Delta^T_l = \left(-\frac16 \dot h + \dot \kappa \Delta^T_0 \right) \delta_{l0} - \frac15 \left(\frac13 \dot \alpha + \frac12 \dot\kappa \Pi \right) \delta_{l2} + \frac13 \delta_{l1} \dot\kappa \theta_{\rm b}/k - \dot\kappa \Delta^T_l + \frac{k}{2l+1} \left[l \Delta^T_{l-1}-(l+1) \Delta^T_{l+1} \right].
\end{equation}
Identifying $\delta_\gamma = 4 \Delta^T_0$ and $\theta_\gamma =  3 k \Delta^T_1$, the $l=0$ equation becomes $\dot \delta_\gamma = -(2/3) \dot h -(4/3) \theta_\gamma$, which agrees with Eq.~(2.4a) in \cite{Blas:2011rf} and Eq.~(10) in \cite{Ma:1995ey}.
If we then set $\sigma_\gamma = 2 \Delta^T_2$, the $l=1$ equation becomes
$\dot\theta_\gamma = -\dot\kappa (\theta_\gamma-\theta_{\rm b}) + k^2 \left(  \dot\delta_\gamma/4 - \sigma_\gamma\right)$,
which agrees with Eq.~(2.4b) in \cite{Blas:2011rf} by identifying our $\dot\kappa$ with their $\tau_c^{-1}$ and $\Theta_{\gamma {\rm b}} = \theta_\gamma-\theta_{\rm b}$.  Our $l=2$ equation is then,
\begin{equation}
     2\dot\sigma_\gamma = \frac{8}{15} \theta_\gamma- \frac35 k F_{\gamma3}+\dot\kappa \left( -\frac25 \Pi-2 \sigma_\gamma\right) - \frac{4}{15} \dot \alpha,
\end{equation}
which agrees with Eq.~(2.4c) by identifying $-\dot\alpha = \dot h + 6\dot\eta$ and $F_{\gamma l} = 4 \Delta^T_l$.  Agreement with their results also requires $-\Pi = (2\sigma_\gamma + G_{\gamma0}+G_{\gamma 2})/4$.  This is consistent with $\Pi = -\Delta^T_2 + \sqrt{6} \Delta^E_2$ if $(G_{\gamma0} + G_{\gamma_2}) = -4\sqrt{6} \Delta^E_2$.

\subsubsection{Polarization}

To be complete, we write the Boltzmann equations for the polarization including the effects of cosmic birefringence.  The equations for no birefringence are recovered by setting $\dot\beta=0$.
We re-write Eq.~(\ref{eqn:EBklm}) in terms of $E_l(\tau) + i B_l(\tau) \equiv (4/3)\sqrt{(l-2)!/(l+2)!} [\Delta^E_l(\tau) + i \Delta^B_l(\tau)]$, in which case Eq.~(\ref{eqn:EBklm}) becomes
\begin{equation}
     E_{l}(\tau) + i B_l(\tau) =  \int_{\tau_i}^\tau\,
     d\tau'\, e^{-\kappa(\tau,\tau')+2i\beta(\tau,\tau')} \dot\kappa (-\Delta^T_2+ 9 E_2) \frac{j_l(w)}{w^2}.
\end{equation}
Using ${\rm Lim}_{w\to0} j_l(w)/w^2 = (1/15) \delta_{l2}$ and also using $j_l(w)/w = \left[j_{l-1}(w) + j_{l+1}(w) \right]/(2l+1)$ and $(d/dw) \left[j_l(w)/w^2 \right] = j_l'(w)/w^2 -(2/w) j_l(w)/w^2$, and separating the real and imaginary parts, we obtain
\begin{eqnarray}
     \dot E_l &=& -\dot \kappa E_l -2 \dot\beta B_l + \frac{1}{15}\dot\kappa (-\Delta^T_2 + 9 E_2) \delta_{l2} + \frac{k}{2l+1}\left[(l-2) E_{l-1} - (l+3)E_{l+1}\right], \nonumber \\
     \dot B_l &=& -\dot \kappa B_l + 2 \dot\beta E_l + \frac{k}{2l+1}\left[(l-2) B_{l-1} - (l+3)B_{l+1}\right].
\end{eqnarray}
If we set $\dot\beta=0$ and take into account that their $E^{(0)}_l$ is $(2l+1) \Delta^E_l$, the above equations agree with Eq.~(63) in \cite{Hu:1997hp}.

\subsection{Tensor equations}

\subsubsection{Temperature fluctuations}

We differentiate Eqs.~(\ref{eqn:TensorDeltaT}) and use ${\rm Lim}_{w\to0} R^{L,TE}_l(w)=- ( \sqrt{2}/5 )\delta_{l2}$  to obtain,
\begin{equation}
      \dot \Delta^T_l = -\dot\kappa \Delta^T_l + \frac{k}{2l+1} \left( \sqrt{l^2-4} \Delta^T_{l-1} - \sqrt{(l+1)^2-4} \Delta^T_{l+1} \right) - \frac{1}{5\sqrt{3}} \dot h^E +\frac{1}{10} \dot\kappa \left( \Delta^T_{2,k} -\sqrt{6} \Delta^E_{2,k}\right) \delta_{l2}.
\end{equation}

\subsubsection{Polarization}

Just as the transformation from integral equation to Boltzmann hierarchy couples modes of different $l$, it also (for tensor metric perturbations) also couples E and B modes.  This happens because the separation of E and B TAM modes is not preserved if the origin is changed.  The polarization Boltzmann equations are derived using,
\begin{eqnarray}
     \frac{d R^{E,E}_l(x)}{dx} & =& \frac{(l-2)(l+2)}{l(2l+1)}R^{E,E}_{l-1}(x)-\frac{(l-1)(l+3)}{(l+1)(2l+1)} R^{E,E}_{l+1}(x) + \frac{4i}{l(l+1)}R^{B,B}_l(x), \nonumber \\
     \frac{d R^{B,B}_l}{dx} &=&\frac{(l-2)(l+2)}{l(2l+1)} R^{B,B}_{l-1}(x) -\frac{(l-1)(l+3)}{(l+1)(2l+1)} R^{B,B}_{l+1}+\frac{4i}{l(l+1)} R^{E,E}_l(x).
\end{eqnarray}
We then differentiate Eqs.~(\ref{eqn:TensorDeltaE}) and (\ref{eqn:DeltaB})} and use ${\rm Lim}_{w\to0} R^{E,E}_l(w)=-(2/5)\delta_{l2}$  and ${\rm Lim}_{w\to0} R^{B,B}_l(w)=0$ to obtain
\begin{eqnarray}
    \dot \Delta^E_l &=& - \dot \kappa \Delta^E_l + \frac{(l-2)(l+2)}{l(2l+1)} \Delta^E_{l-1}(x)-\frac{(l-1)(l+3)}{(l+1)(2l+1)} \Delta^E_{l+1}(x) + \frac{4i}{l(l+1)}\Delta^B_l(x) - \frac{\sqrt{6}}{10}\dot \kappa \left( \Delta^T_{2,k} -\sqrt{6} \Delta^E_{2,k}\right) \delta_{l2}, \nonumber \\
    \dot \Delta^B_l &=& -\dot\kappa \Delta^B_l + \frac{(l-2)(l+2)}{l(2l+1)} \Delta^B_{l-1}(x) -\frac{(l-1)(l+3)}{(l+1)(2l+1)} \Delta^B_{l+1}+\frac{4i}{l(l+1)} \Delta^E_l(x),
\end{eqnarray}
as derived first in \cite{Hu:1997hp}.\footnote{The correspondence with their Eqs.~(60), (63), and (64) is obtained by noting that their moments differ by ours by a factor of $2l+1$.  It also requires that we identify our $\dot h^E$ with their $\sqrt{3} \dot H$, which follows from the definition in their Eq.~(34) and (38) of $H$.}

\section{Anisotropic cosmic birefringence}
\label{app:aniso}

Section~\ref{sec:cb} assumed a spatially uniform rotation, but it is also conceivable that $\beta$ varies as a function of position on the sky \citep{Kamionkowski:2008fp,Gluscevic:2009mm,Caldwell:2011pu,Yadav:2009eb}, and equations for the evolution of photon temperature and polarization in the presence of anisotropic $\beta$ were derived recently in \cite{Greco:2024oie,Namikawa:2024dgj}.  Here we derive the polarization induced by anisotropic rotation, working only to leading order in the rotation angle, where the results are reasonably simple.  The derivation may also provide additional insights on the uniform-birefringence result.  We leave the calculation beyond linear order for future work.

We will start by considering a single TAM wave of quantum numbers $klm$ for the metric perturbation and suppose it is subjected to a rotation angle $\dot\beta(\tau)= \dot\beta_{LM}(\tau) Y_{LM}(\hat n)$.  We will find that at linear order in $\beta_{LM}$ there are B-mode polarization modes of quantum numbers $l'm'$ induced with $|L-l| \leq l' \leq |L+l'|$.  To begin, we infer from Eq.~(\ref{eqn:Eklm}) that the polarization pattern induced by a single TAM wave is
\begin{equation}
    {\cal P}_{ab}(\vec x,\hat n,\tau) = -\frac{\sqrt{3}}{2} \int_{\tau_i}^\tau \, d\tau'\, e^{-\kappa(\tau,\tau')} \dot \kappa \Pi_{klm}(\tau') R^{L,TE}_l(w) Y_{(lm)ab}^E(\hat n).
\end{equation}
With the given $LM$ mode for cosmic birefringence, this gets changed, to linear order in $\beta_{LM}$, to
\begin{equation}
    {\cal P}_{ab}(\vec x,\hat n,\tau) = -\frac{\sqrt{3}}{2} \int_{\tau_i}^\tau \, d\tau'\, e^{-\kappa(\tau,\tau')} \dot \kappa \Pi_{klm}(\tau') R^{L,TE}_l(w) \left[ Y_{(lm)ab}^E(\hat n) + 2\beta_{LM} Y_{LM}(\hat n) Y_{(lm)ab}^B(\hat n) \right].
\end{equation}
We then use Eqs.~(65)--(66) of \cite{Dai:2012ma}, which we write as
\begin{equation}
  \int d^2\hat n\,\big[Y^{E}_{(l'm')ab}(\hat n) +iY^{B}_{(l'm')ab}(\hat n)\big]^{*}
  Y_{(LM)}(\hat n)\,Y^{B\,ab}_{(lm)}(\hat n)
  = i\,\mathcal{C}^{\,m'mM}_{\,l'\,l\,L},
  \label{eq:overlap}
\end{equation}
with
\begin{equation}
  \mathcal{C}^{\,m'mM}_{\,l'lL}
  \equiv (-1)^{m'}\,G^{\,l'\,l\,L}_{-m'\,m\,M}\,
  \frac{\langle l'2\,l,-2|L0\rangle}{\langle l'0\,l0|L0\rangle},
  \label{eq:coupling}
\end{equation}
where $G^{l_1l_2l_3}_{m_1m_2m_3}$ is the Gaunt integral [Eq.~(36) of \cite{Dai:2012ma}].   We then find that this combination of a $klm$ density mode and an $LM$ CB mode induces corrections
\begin{equation}
    \delta (E_{l'm'} + i  B_{l'm'})_{klm,LM} = -\sqrt{3} i \mathcal{C}^{m'mM}_{l'lL} \int_{\tau_i}^\tau\, d\tau'\, e^{-\kappa(\tau,\tau')} \dot \kappa \Pi_{klm}(\tau') R^{L,TE}_l(w) \beta_{LM}(\tau,\tau'),
\end{equation}
to all $l'm'$ polarization moments that satisfy $m=m'+M$ and $|l-L|\le l'\le l+L$.  Thus, to linear order in $\beta_{LM}$, the net effect is a change,
\begin{equation}
    \delta (E_{l'm'} + i B_{l'm'})= -\sqrt{3} i \sum_k \sum_{|l-L|\le l'\le l+L} \sum_{M=m-m'}  \mathcal{C}^{m'mM}_{l'lL} \int_{\tau_i}^\tau\, d\tau'\, e^{-\kappa(\tau,\tau')} \dot \kappa \Pi_{klm}(\tau') R^{L,TE}_l(w) \beta_{LM}(\tau,\tau').
\end{equation}
To go beyond linear order becomes far more complicated.  The result at linear order in $\beta_{LM}$ is obtained by taking the sources $\Pi_{klm}(\tau)$ to zero-th order.  To go beyond this linear calculation, though, the corrections to $\Pi_{klm}(\tau)$ need to be included, but each $\Pi_{klm}(\tau)$ then receives contributions from all $kl'm'/LM$ combinations that satisfies the Gaunt conditions and the evolution of all the different $klm$ modes thus becomes coupled.

\bibliographystyle{elsarticle-harv} 
\bibliography{refs.bib}

\end{document}